\documentclass[aps,prl,floatfix,epsfig,twocolumn,showpacs,preprintnumbers]{revtex4}
\usepackage{amssymb}
\usepackage{graphicx}
\usepackage{amsmath}

\begin{document}

\title{Density-functional calculations of the electronic structure and
lattice dynamics of superconducting LaO$_{0.5}$F$_{0.5}$BiS$_{2}$: Evidence
for an electron-phonon interaction near the charge-density-wave instability}
\author{Xiangang Wan,$^{1,3}$ Hang-Chen Ding,$^{2}$ Sergey Y. Savrasov$^{3}$
and Chun-Gang Duan$^{2,4}$}
\affiliation{$^{1}$National Laboratory of Solid State Microstructures and Department of
Physics, Nanjing University, Nanjing 210093, China\\
$^{2}$Key Laboratory of Polar Materials and Devices, Ministry of Education,
East China Normal University, Shanghai 200062, China \\
$^{3}$Department of Physics, University of California, Davis, One Shields
Avenue, Davis, CA 95616, USA \\
$^{4}$National Laboratory for Infrared Physics, Chinese Academy of Sciences,
Shanghai 200083, China}
\date{\today }

\begin{abstract}
We discuss the electronic structure, lattice dynamics and electron--phonon
interaction of newly discovered superconductor LaO$_{0.5}$F$_{0.5}$BiS$_{2}$
using density functional based calculations. A strong Fermi surface nesting
at $\mathbf{k}$=($\pi $,$\pi $,0) suggests a proximity to charge density
wave instability and leads to imaginary harmonic phonons at this $\mathbf{k}$
point associated with in--plane displacements of S atoms. Total energy
analysis resolves only a shallow double-well potential well preventing the
appearance of static long--range order. Both harmonic and anharmonic
contributions to electron--phonon coupling are evaluated and give a total
coupling constant $\lambda \simeq 0.85$ prompting this material to be a
conventional superconductor contrary to structurally similar FeAs materials.
\end{abstract}

\date{\today }
\maketitle

Superconductors with layered crystal structures such as cuprates\cite{HTC},
ruthenates\cite{Sr2RuO4}, or MgB$_{2}$\cite{MgB2} have generated enormous
research interest. A recent discovery of iron pnictides\cite{Fe-based} have
triggered another wave of extensive studies \cite{Fe-based-2}, and while the
mediator of pairing in these systems remains officially unidentified, a
large amount of evidence points to magnetic spin fluctuations\ induced by
antiferromagnetic spin--density--wave (SDW) instability due to
Fermi--surface nesting\cite{SDW} at wave vector $\mathbf{k}$=($\pi ,\pi ,0)$
similar to the cuprates. Usually changing the blocking layer can tune the
superconducting T$_{c}$, thus searching for new layered superconductors is
of both fundamental and technological importance.

Very recently, a new layered superconductor Bi$_{4}$O$_{4}$S$_{3}$ has been
found\cite{Bi4O4S3}, and soon after, two other systems, LaO$_{1-x}$F$_{x}$BiS%
$_{2}$ \cite{LaOFBiS2}and NdOBiS$_{2}$ \cite{NdOBiS2} have been discovered.
Here, the basic structural unit is the BiS$_{2}$ layer which is similar to
the Cu--O planes in Cu--based superconductors\cite{HTC} and the Fe--A (A=P,
As, Se, Te) planes in iron pnictides\cite{Fe-based-2}. A chance to explore
superconductivity and increase T$_{c}$ in these new compounds has already
resulted in a lot of work that appeared shortly after the discovery\cite%
{Bi4O4S3,LaOFBiS2,NdOBiS2,Bi4O4S3-2,Bi4O4S3-3,Pressure,Minimal electronic
models,SpinExc,LaO0.5F0.5BiS2,Singh}. Hall effect measurements reveal
multiband features and suggest the superconducting pairing occurs in
one--dimensional chains \cite{Bi4O4S3-2}. It was proposed that these
compounds are type II superconductors and good candidates for thermoelectric
materials \cite{Bi4O4S3-3}. Electrical resistivity measurements under
pressure reveal that Bi$_{4}$O$_{4}$S$_{3}$ and La(O,F)BiS$_{2}$ have
different T$_{c}$ versus pressure behavior, and the Fermi surface of
La(O,F)BiS$_{2}$ may be located in the vicinity of some instabilities\cite%
{Pressure}. A two \textit{p} bands electronic model has been proposed based
on band structure calculation\cite{Minimal electronic models}, and a good
Fermi--surface nesting with wave vector $\mathbf{k}$=($\pi ,\pi ,0)$ has
been found\cite{Minimal electronic models}. The importance of the nesting
has been emphasized experimentally\cite{Pressure} and it was suggested that
electronic correlations may play a role in superconductivity of these
systems \cite{SpinExc}.

\begin{figure}[tbp]
\includegraphics[width=3.8in]{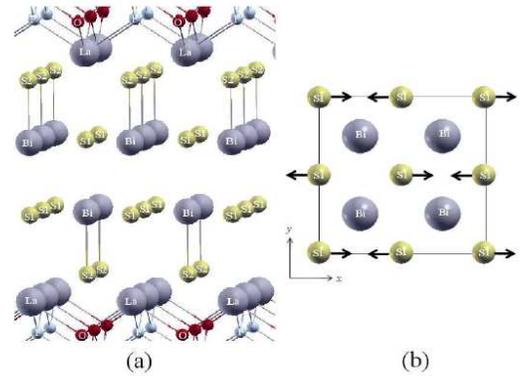}
\caption{(a) Structure of La(O$_{0.5}$F$_{0.5}$)BiS$_{2}$; (b) unstable
phonon mode corresponding to wave vector k=($\protect\pi $,$\protect\pi $%
,0). }
\end{figure}

Here we report our theoretical studies of the electronic structure and
lattice dynamic properties for LaO$_{0.5}$F$_{0.5}$BiS$_{2}$, the compound
that posses the highest T$_{c}\simeq $10 K \cite{LaOFBiS2} among known BiS
based materials\cite{Bi4O4S3,LaOFBiS2,NdOBiS2} and whose structure is
similar to superconducting iron arsenides LaFeO$_{1-x}$F$_{x}$As\cite%
{Fe-based}. Our first--principles calculations are based on density
functional theory (DFT) and show that the bands around the Fermi level are
not sensitive to F substitution, thus making rigid band approximation
adequate for LaO$_{1-x}$F$_{x}$BiS$_{2}$. We find a strong nesting of the
Fermi surface at $\mathbf{k}=$($\pi ,\pi ,0)$ for x=0.5 by performing the
calculation for ordered compound LaO$_{0.5}$F$_{0.5}$BiS$_{2}$. Our linear
response based phonon calculation shows that the nesting results in a large
phonon softening at this $\mathbf{k}$ and in the appearance of the imaginary
modes associated with in--plane displacements of S atoms. Our $\sqrt{2}%
\times \sqrt{2}\times 1\ $supercell total--energy calculation finds a
shallow double--well potential prompting that these displacements are
dynamic. Contrary to the expectations that electronic correlations may play
a role in these systems, our calculated electron--phonon coupling constant $%
\lambda \simeq 0.85$ suggests that this material is a conventional
superconductor. However, our anharmonic model calculation shows that the
vicinity of the charge--density--wave (CDW) instability is essential for the
superconductivity which is reminiscent to iron pnictides whose proximity to
SDW is well established\cite{Fe-based}.

\begin{figure}[tbp]
\includegraphics[width=3.5in]{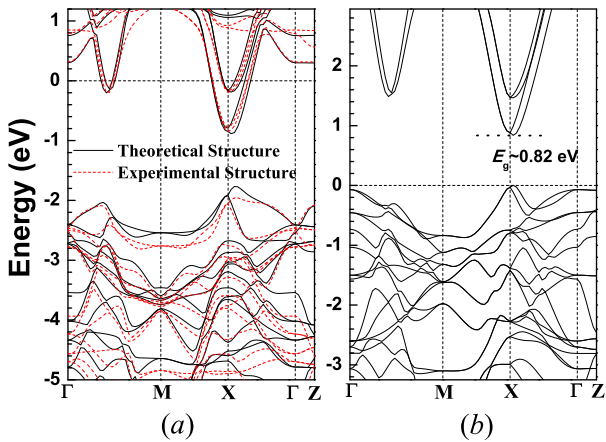}
\caption{Band structure of (a) LaO$_{0.5}$F$_{0.5}$BiS$_{2}$; (b) LaOBiS$%
_{2} $.}
\end{figure}

Our electronic structure calculations are performed within the
generalized gradient approximation (GGA)\cite{GGA}. Whenever
possible, we cross--check the results by two different commonly
used total--energy codes: a) the Vienna Ab--Initio Simulation
Package (VASP)\cite{VASP} and b) the Quantum ESPRESSO package
(QE)\cite{pwscf}. The consistency of our results for two sets of
calculations is satisfactory. We use a 500 eV plane--wave cutoff
and a dense 18$\times $ 18$\times $ 6 k--point mesh in the
irreducible Brillouin zone (IBZ) for self--consistent
calculations. For structural optimization, the positions of ions
were relaxed towards equilibrium until the Hellman--Feynman forces
became less than 2 meV/\r{A}. For the phonon calculations, we
adopt a scalar relativistic version of density--functional
linear--response method\cite{Linear-response} as implemented in QE\cite%
{pwscf}. For consistency, the results presented in this paper are
obtained by scalar relativistic version of QE, unless otherwise
specified. The effect of spin-orbit coupling (SOC) is discussed in
the Supplemental Material.

Powder x--ray diffraction (XRD) pattern shows that LaO$_{1-x}$F$_{x}$BiS$_{2}
$ forms a layered crystal structure with a space group P4/nmm\cite{LaOFBiS2}%
. La, S and Bi locate at 2\emph{b} position, while O/F take the 2\emph{a}
site. Similar to LaO$_{1-x}$F$_{x}$FeAs\cite{Fe-based}, the structure
consists of alternating La(O$_{1-x}$F$_{x}$) and BiS$_{2}$ layers\cite%
{LaOFBiS2}. One BiS$_{2}$ layer contains two BiS planes (namely Bi-S1 plane)
and two pure S planes (i.e. S2 plane) as shown in Fig.1a. To study the
influence of F doping, we carried out calculations for two systems, i.e.
LaOBiS$_{2}$ (x=0) and LaO$_{0.5}$F$_{0.5}$BiS$_{2}$ (x=0.5). Being embedded
into LaO plane, we expect that the substitution by F has only small effect
on the BiS$_{2}$ layer. Thus we simulate LaO$_{0.5}$F$_{0.5}$BiS$_{2}$ by
replacing half of the Oxygen 2\textit{a}-sites by F\ orderly, despite the
substitution may be random in reality. The optimized lattice parameters and
Wyckoff positions for each atom are shown in Table I, together with
available experimental data \cite{LaOFBiS2}. While the overall agreement
between numerical and experimental structures is good, Table I reveals an
interesting aspect: the differences between the experimental and theoretical
values of the \textbf{\emph{z}}-coordinate of S1 are unusually large,
similar as in the FeAs-superconductors\cite{Yin-Pickett}. There is also a
large difference between the numerical and experimental inter-layer
distance. Note that we have also performed the internal atomic coordinates
optimization based on the experimental lattice parameters (shown in the
Table 1S of Supplementary Information). As can be seen from Table 1S of
Supplementary Information, the lattice parameters, namely the inter-layer
distance has only small effect on the \textbf{\emph{z}}-coordinate of S1.\
In Table I we also list the numerical data for LaOBiS$_{2}$. Comparing with
the results for LaO$_{0.5}$Bi$_{0.5}$S$_{2}$, one can conclude that F
substitution has only little effect on the lattice parameters, and a small
effect on the BiS$_{2}$ layer.

\begin{table}[tbp]
\caption{Calculated lattice parameters and Wyckoff positions of LaOBiS$_{2}$
and La(O$_{0.5}$F$_{0.5}$)BiS$_{2}$. Experimental results \protect\cite%
{LaOFBiS2} are also listed for comparison.}%
\begin{tabular}{llllll}
\hline
&  & LaOBiS$_{2}$ &  & La(O$_{0.5}$F$_{0.5}$)BiS$_{2}$ &  \\ \hline
& Site & Cal. & Exp. & Cal. & Exp. \\ \hline
\emph{a}(\AA ) &  & 4.0394 & -- & 4.0780 & 4.0527 \\
\emph{c}(\AA ) &  & 14.1232 & -- & 13.4925 & 13.3237 \\
\emph{z} & La (2\emph{b}) & 0.0889 & -- & 0.1049 & 0.1015 \\
\emph{z} & Bi (2\emph{b}) & 0.6304 & -- & 0.6141 & 0.6231 \\
\emph{z} & S1 (2\emph{b}) & 0.3932 & -- & 0.3902 & 0.3657 \\
\emph{z} & S2 (2\emph{b}) & 0.8090 & -- & 0.8134 & 0.8198 \\
\emph{z} & O (2\emph{a}) & 0.0000 & -- & 0.0000 & 0.0000 \\
\emph{z} & F (2\emph{a}) & -- & -- & 0.0000 & 0.0000 \\ \hline
\end{tabular}%
\end{table}

The discrepancy between the experimental and theoretical positions of S1
atoms results in a considerable difference for valence band as shown in
Fig.2a. We also perform calculation based on the numerical internal atomic
coordinates and experimental lattice parameters. Comparing with the results
based on theoretical structure shows that changing inter-layer distance only
slightly affects the band structure, and the considerable difference shown
in Fig.2a is mainly due to \textbf{\emph{z}}-coordinate of S1. For LaOBiS$%
_{2}$ both S 3\textit{p} and O 2\textit{p} states appear mainly between -4.0
and 0.0 eV. Although located\ primarily above the Fermi level, Bi 6\textit{p}
have also a considerable contribution to the states between -4.0 and 0.0 eV,
indicating\ a strong hybridization between Bi 6\textit{p} and S 3\textit{p}
states. In agreement with previous calculation\cite{Minimal electronic
models} our results show that LaOBiS$_{2}$ is an insulator with a band gap
of 0.82 eV.\textbf{\ }A little dispersion along $\Gamma $\ to Z line clearly
shows a two dimensional character of the band structure which indicates that
the interlayer hybridization is small. Comparing Fig.2a and Fig.2b, it is
clear that the main influence of F substitution\ is a carrier doping
characterized by the associated upshift of the Fermi level towards the Bi 6%
\textit{p} band, and the system becomes metallic as shown in Fig.2a. It is
interesting to notice that doping by F has a negligible effect on the lowest
conduction band. Thus we conclude that rigid band approximation is valid for
LaO$_{1-x}$F$_{x}$BiS$_{2}$.

\begin{figure}[tbp]
\includegraphics[width=3in]{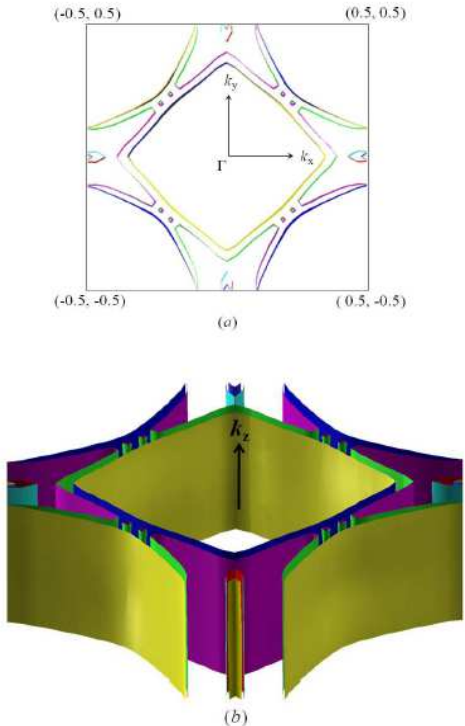} 
\caption{Calculated Fermi surface of La(O$_{0.5}$F$_{0.5}$)BiS$_{2}$: (a)
cross section for k$_{z}$=0, (b) 3D view.}
\label{Fig-FS}
\end{figure}

There are four bands that cross the Fermi level and result in a large two
dimensional--like Fermi surface that is shown in Fig. 3. Consistent with the
tight--binding result\cite{Minimal electronic models}, our density
functional calculation shows a strong Fermi surface nesting at wavevectors
near $\mathbf{k}=(\pi $,$\pi ,0)$.

The Bi 6\textit{p} orbitals are spatially extended and strongly hybridized
with S 3\textit{p} states near the Fermi energy. Thus we do not expect
electronic correlations to be essential for this compound. To check whether
the conventional electron--phonon mechanism can be responsible for
superconductivity here, we first perform a linear response phonon
calculation \cite{Linear-response} as implemented in QE\cite{pwscf}. An 18$%
\times $18$\times $6 grid was\ used for the integration over IBZ. Our
calculated phonon spectrum along major high symmetry lines of the Brillouin
zone is shown in Fig.4 where the phonon dispersions are seen to extend up to
400 cm$^{-1}$. The phonon modes have only a little dispersion along $\Gamma $%
-Z direction, which again indicates the smallness of the interlayer
coupling. There are basically three panels in the phonon spectrum that are
easily distinguished along the $\Gamma $-Z direction. The top four branches
above 300 cm$^{-1}$ are mainly contributed by O and F, while the branches
below 80 cm$^{-1}$ come from the BiS$_{2}$ layer. The phonon vibrations
within the \textit{xy}--plane show a significant dispersion as shown in
Fig.4. Analyzing the evolution of the phonon eigenvectors in the Brillouin
zone reveals that there is clear separation between the \textit{xy} and
\textit{z} polarized vibrations, and most of the modes show a definite
in-plane or out-of-plane character.

\begin{figure}[tbp]
\includegraphics[width=3.0in]{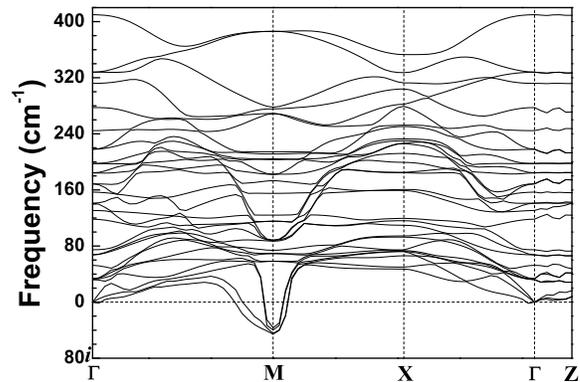}
\caption{Calculated phonon dispersions for LaO$_{0.5}$F$_{0.5}$BiS$_{2}$
using density functional linear response approach.}
\end{figure}

A striking feature of this phonon spectrum is the presence of phonon
softening around the M point that we associate with the strong Fermi surface
nesting. We find that there are four totally unstable modes, mainly
contributed by the S1 in--plane vibrations, where they either displace along
x or y direction, and either in--phase or out-of--phase between the two BiS
planes. The polarization vectors are shown in Fig.1b.
\begin{figure}[tbp]
\includegraphics[width=3.0in]{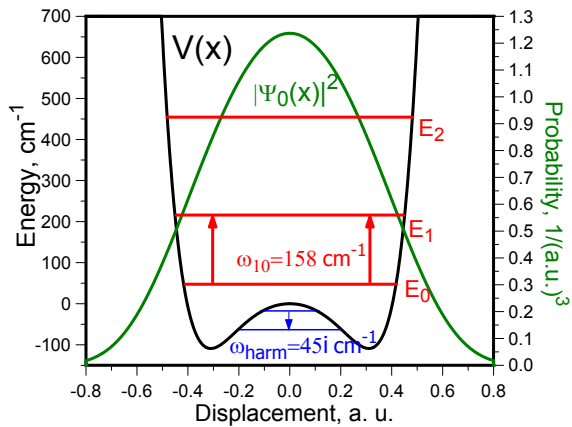}
\caption{Calculated double well potential for the unstable phonon mode using
the frozen phonon method and its first three eigenstates.}
\end{figure}

To understand whether the CDW instability is present in this material, we
perform a frozen phonon analysis with the unit cell doubled ($\sqrt{2}\times
\sqrt{2}\times 1$) according to the ($\pi $,$\pi $,0) nesting wave vector.
We perform four calculations by moving the atoms according to the
eigenvectors of the four unstable phonon modes at the M point. Our frozen
phonon calculations show a shallow double well potential where the S1 atoms
shift about 0.18 \r{A}\ away from the original high symmetry position as we
show in Fig.5. To illustrate the crucial change in the electronic structure
due to M\ point frozen phonon S1 in-plane motion, we show in Fig. 6 the band
structures of distorted and undistorted structures in the vicinity of the
Fermi level. The black lines depict the undistorted energy bands while the
red lines correspond to the frozen-phonon distorted structure with S1
displacement by 0.18 \r{A}, For comparison, both lines are drawn in an $%
\sqrt{2}\times \sqrt{2}\times 1$\ supercell. A considerable difference in
band structures induced by the in--plane S1 displacement indicates a large
electron--phonon coupling from this CDW instability.

\begin{figure}[tbp]
\includegraphics[width=3in]{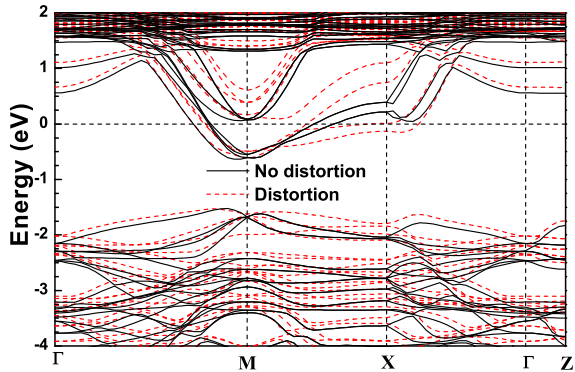}
\caption{Calculated band structure. Black line is the undistorted bands, red
line is a distorted bands corresponding to the \ S1 in-plane displacement.
See text for details.}
\end{figure}

The depth of the double well is only about 100 cm$^{-1}$ indicates that the
displacements are dynamic. By solving a corresponding anharmonic oscillator
problem we deduce a ground state atomic wave function which is indeed
centered at the high symmetry position as shown in Fig. 5 together with the
first three energy levels. It is therefore clear that our instable modes are
not related to a statically distorted structure of LaO$_{0.5}$F$_{0.5}$BiS$%
_{2}.$\ This is consistent with experimental observations that the
resistivity changes smoothly from 300 K to about 10 K where the
superconductivity occurs\cite{LaOFBiS2,Pressure}.

We finally turn our discussion to the wave--vector ($\mathbf{q}$) and mode ($%
\nu )$ dependent electron--phonon coupling $\lambda _{\nu }(\mathbf{q})$. At
first, this can be done for all stable phonons. Calculation shows that the
O/F modes have negligible contribution to electron-phonon coupling. With
strong hybridization, the coupling, however, is relatively strong for the
BiS based modes. For example, we can find $\lambda $'s of the order of 1 for
the S2 based optical phonons around 310 cm$^{-1}$ near the $\Gamma $\ point.
The analysis of the polarization vectors shows that these vibrations involve
S2 movements towards Bi atoms. Unfortunately, finding the integral value of $%
\lambda $ is a challenging problem due to the appearance of the imaginary
frequencies, although neglecting completely the unstable modes results in
already large average coupling constant ($0.75$) calculated using 4$\times $4%
$\times $2\ q-mesh. This is mainly due to the discussed S1/S2 vibrations.

To find the contribution for the four anharmonic modes, we follow the
strategy of Ref. \cite{Hui}, where transitions from the ground to all phonon
excited states of the anharmonic well need to be taken into account. The
detailed theory using total energy frozen phonon method has been elaborated
in Ref. \cite{Meregalli}. Our numerical value of $\lambda $ for the four
anharmonic modes at the M point is 0.4. It has to be weighted somewhat by
the area of the Brillouin Zone where the actual instability occurs therefore
adding it the result for harmonic $\lambda \ $\ should give us a total
coupling constant of 0.85. Inserting this value into the McMillan formula
for T$_{c}$, with the Coulomb parameter $\mu ^{\ast }\simeq 0.1$\ and $%
\omega _{D}=260K$ yields values of T$_{c}\simeq 11.3\ K$ in reasonable
agreement with experiment.

Finally, we have also performed calculations to check the effect
of SOC, and show the results in the Supplemental Material. As can
be seen from the Supplemental Material, the optimization including
SOC gives very similar lattice parameters and internal atomic
coordinates as those without SOC. We find that SOC does affect the
band structure a little, but it does not change the Fermi surface.
As shown in the Fig.S2 of Supplemental Material, the Fermi surface
nesting does not disappear in the presence of SOC. Moreover, the
CDW instability also remains to be active, as shown in the Fig.S3
of Supplemental Material. The band calculation with SOC also shows
that S1 in-plane motion has a large effect on the band structure,
same as the case without SOC (Fig.\ S4 of Supplemental Material),
indicating strong electron-phonon coupling. The large
electron-phonon coupling can be also found from the comparison of
band structure from theoretical structure and experimental
structure (Fig. 5S of Supplemental Material).

In conclusion, electronic structure, lattice dynamics and electron--phonon
interaction of the newly found superconductor LaO$_{0.5}$F$_{0.5}$BiS$_{2}$
have been investigated using density functional theory and linear response
approach. A strong Fermi surface nesting at ($\pi $,$\pi $,0) results in
large phonon softening and strongly anharmonic double well behavior of the
total energy as a function of the in-plane S displacements. A large
electron--phonon coupling constant $\lambda $=0.85 is predicted, which
emphasizes that LaO$_{0.5}$F$_{0.5}$BiS$_{2}$ is a strongly coupled
electron--phonon superconductor.

X.G.W acknowledge useful conversations with Prof. H.H. Wen, J. X. Li and
Q.H. Wang. The work was supported by the National Key Project for Basic
Research of China (Grants No. 2011CB922101 and No. 2010CB923404), NSFC under
Grants No. 91122035, 11174124, 10974082, 61125403 and 50832003), PAPD,
PCSIRT, NCET. Computations were performed at the ECNU computing center.
S.Y.S was supported by by DOE Computational Material Science Network (CMSN)
Grant No. DESC0005468.

\setcounter{figure}{0} \makeatletter
\renewcommand{\thefigure}{S\@arabic\c@figure}

\section{Supplementary Material}

We have optimized all independent internal atomic coordinates
based on the experimental lattice parameter, and show the obtained
results in Table 1S. Note here for consistency, the results shown
in Supplementary Material are all obtained using Vienna Ab--Initio
Simulation Package (VASP). As can be seen from the Table 1S, the
inter-layer distance has only small effect on the discrepancy
between theoretical and experimental \emph{z}-coordinates of S1
atom.

\begin{table}[tbp]
\caption{The numerical Wyckoff positions of
La(O$_{0.5}$F$_{0.5}$)BiS$_{2}$ based on the experimental lattice
parameter. Experimental Wyckoff positions
\protect\cite{LaOFBiS2} are also listed for comparison.}%
\begin{tabular}{llll}
\hline &  & La(O$_{0.5}$F$_{0.5}$)BiS$_{2}$ &  \\ \hline & Site &
Cal. & Exp. \\ \hline
\emph{z} & La (2\emph{b}) & 0.1064 & 0.1015 \\
\emph{z} & Bi (2\emph{b}) & 0.6149 & 0.6231 \\
\emph{z} & S1 (2\emph{b}) & 0.3838 & 0.3657 \\
\emph{z} & S2 (2\emph{b}) & 0.8137 & 0.8198 \\
\emph{z} & O (2\emph{a}) & 0.0000 & 0.0000 \\
\emph{z} & F (2\emph{a}) & 0.0000 & 0.0000 \\ \hline
\end{tabular}%
\end{table}

To study the influence of spin-orbit coupling (SOC), we have also
performed the calculation including SOC. The optimized lattice
parameters and Wyckoff positions of
La(O$_{0.5}$F$_{0.5}$)BiS$_{2}$ are shown in Table 2S. For
comparison, we also list the corresponding values from the
calculation without SOC in Table 2S. It is found that the SOC has
only very small effect on the lattice structure of
La(O$_{0.5}$F$_{0.5}$)BiS$_{2}$. Comparing with the results
without SOC from QE (Table I in the manuscript) shows the
consistency of VASP\ and QE is satisfactory.
\begin{table}[tbp]
\caption{Lattice parameters and Wyckoff positions of La(O$_{0.5}$F$_{0.5}$%
)BiS$_{2}$ from the calculation with and without SOC.}%
\begin{tabular}{llll}
\hline & Site & without SOC & with SOC \\ \hline
\emph{a}(\r{A}) & -- & 4.0782 & 4.0781 \\
\emph{c}(\r{A}) &  & 13.4835 & 13.4566 \\
\emph{z} & La (2\emph{b}) & 0.1055 & 0.1051 \\
\emph{z} & Bi (2\emph{b}) & 0.6161 & 0.6163 \\
\emph{z} & S1 (2\emph{b}) & 0.3891 & 0.3831 \\
\emph{z} & S2 (2\emph{b}) & 0.8154 & 0.8149 \\
\emph{z} & O (2\emph{a}) & 0.0000 & 0.0000 \\
\emph{z} & F (2\emph{a}) & 0.0000 & 0.0000 \\ \hline
\end{tabular}%
\end{table}

We show the band structure with and without the SOC in Fig. 1S. As
can be seen from Fig. 1S, the SOC has a considerable effect on the
bands around $X$ point. To be specific, there are four bands
crossing the Fermi level in the band structure without SOC,
whereas including SOC pushes two bands above the Fermi level.
Nevertheless, it is very interesting to notice that the Fermi
surface from SOC calculation is still quite similar to that
without SOC (see Fig. 2S). As shown in Fig. 2S, including SOC does
not change the two dimensional-like feature of the Fermi surface.
Moreover, the strong Fermi surface nesting can still be found at
wavevectors near $k=$($\pi ,\pi ,0$), as shown in Fig. 2S(b), and
the main effect of SOC is to eliminate a small pocket around X
point.

To check the effect of SOC on the CDW instability, we also perform
four SOC calculations by moving the atoms according to the
eigenvectors of the four unstable phonon modes at the M point. Our
frozen phonon calculations show that the double well potential
remains, though the well-depth drops considerably, as shown in
Fig. 3S.

We show in Fig. 4S the band structures of distorted and
undistorted structures due to M\ point frozen phonon S1 in-plane
motion in the vicinity of the Fermi level. Same with the
calculation without SOC, considerable difference in band
structures induced by the in-plane S1 displacement which again
indicates a large electron-phonon coupling from this CDW
instability. The large electron-phonon coupling can be also found
from the comparison of band structure from theoretical structure
and experimental structure as shown in Fig. 5S.

As a final comment, we recall that the calculation without SOC is
found for heavy element Pb to be satisfactory \cite{Pb} therefore
we do not foresee significant changes in our estimation of the
electron--phonon interaction when the SOC effect is taken into
account.

\begin{figure}[tbp]
\includegraphics [height=3in,width=2.5in] {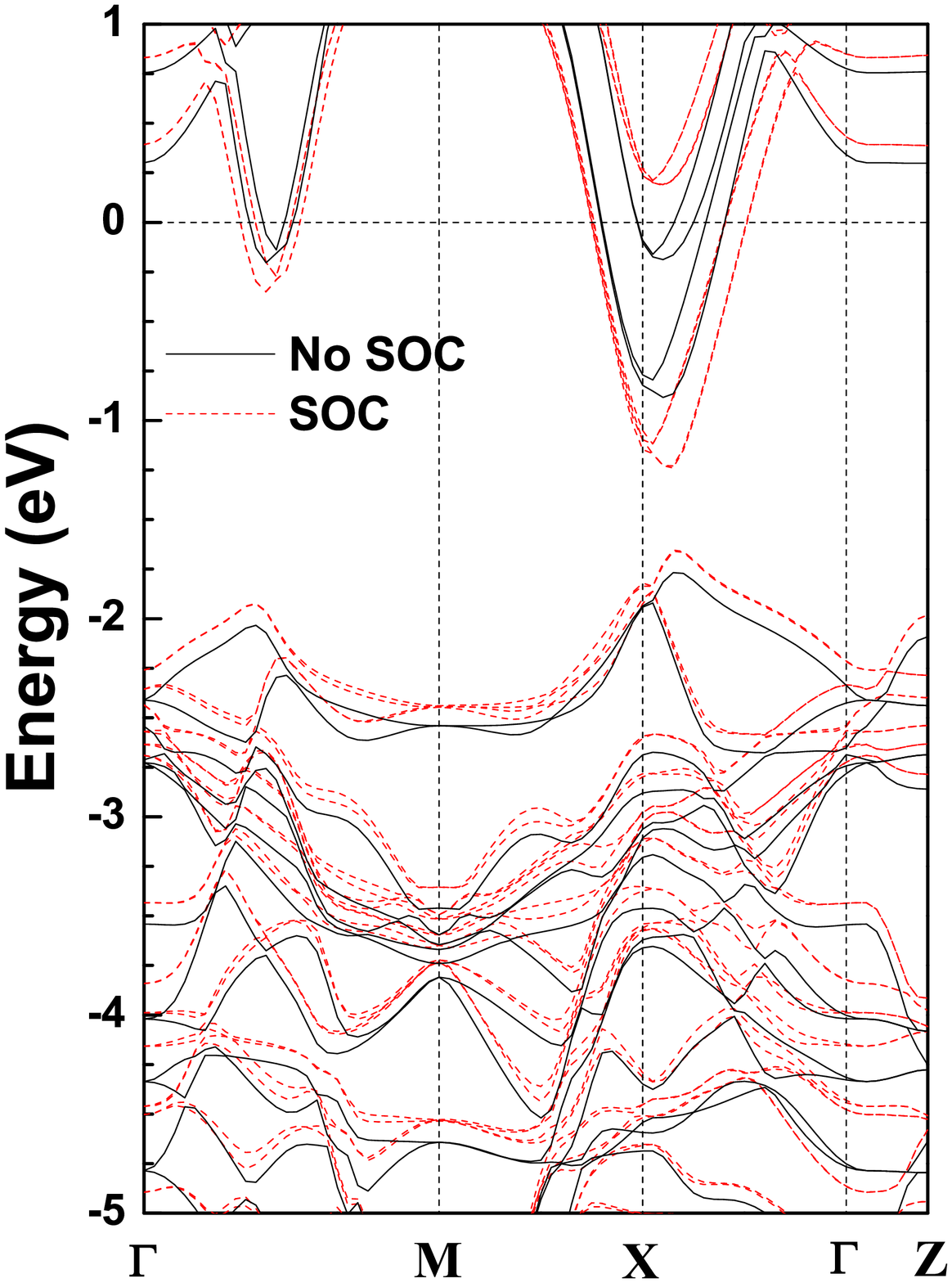}
\caption{Band structure plot}
\end{figure}

\begin{figure}[tbp]
\includegraphics [height=2.5in] {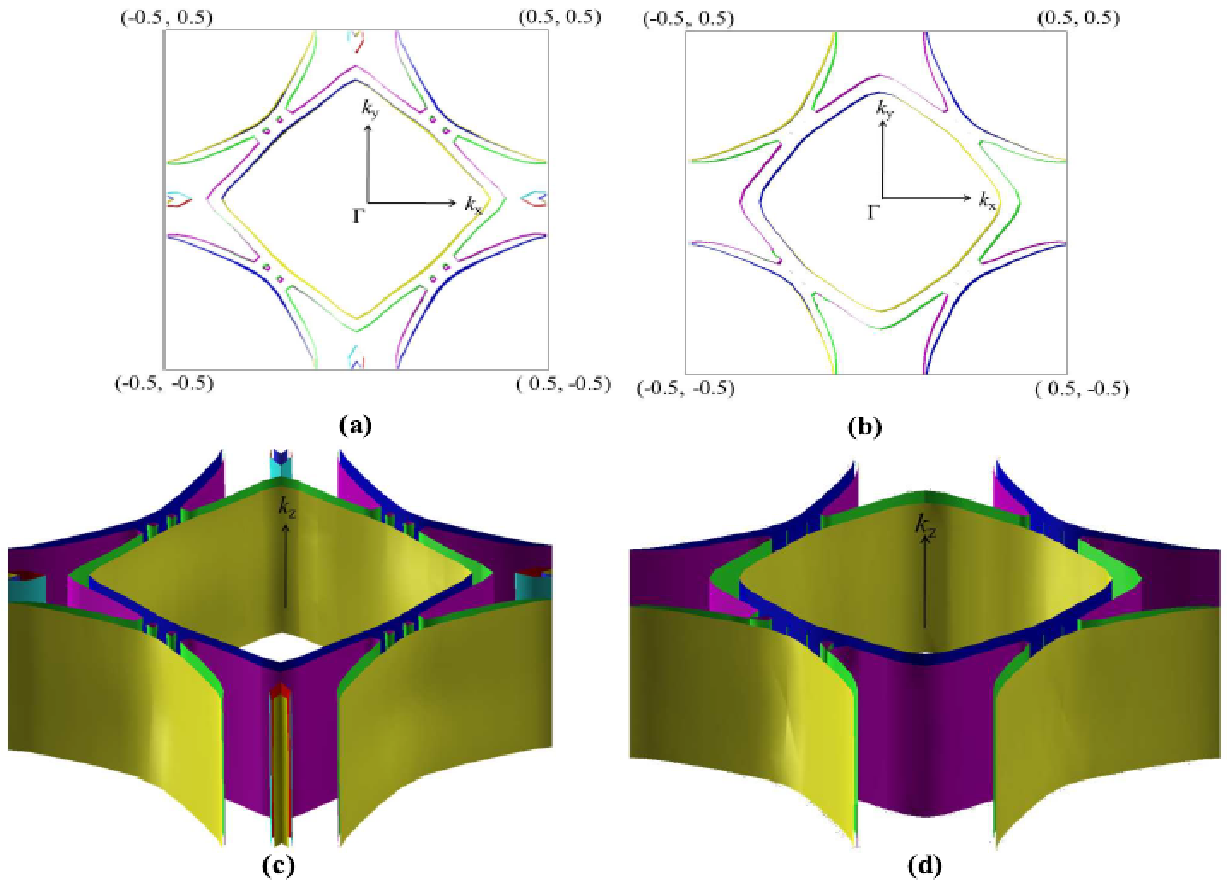}
\caption{Fermi surface of La(O$_{0.5}$F$_{0.5}$)BiS$_{2}$ cross section for k%
$_{z}$=0: (a) without SOC, (b) with SOC and 3D view: (c) without
SOC, (d) with SOC.}
\end{figure}

\begin{figure}[tbp]
\includegraphics [height=2.5in,width=3.0in] {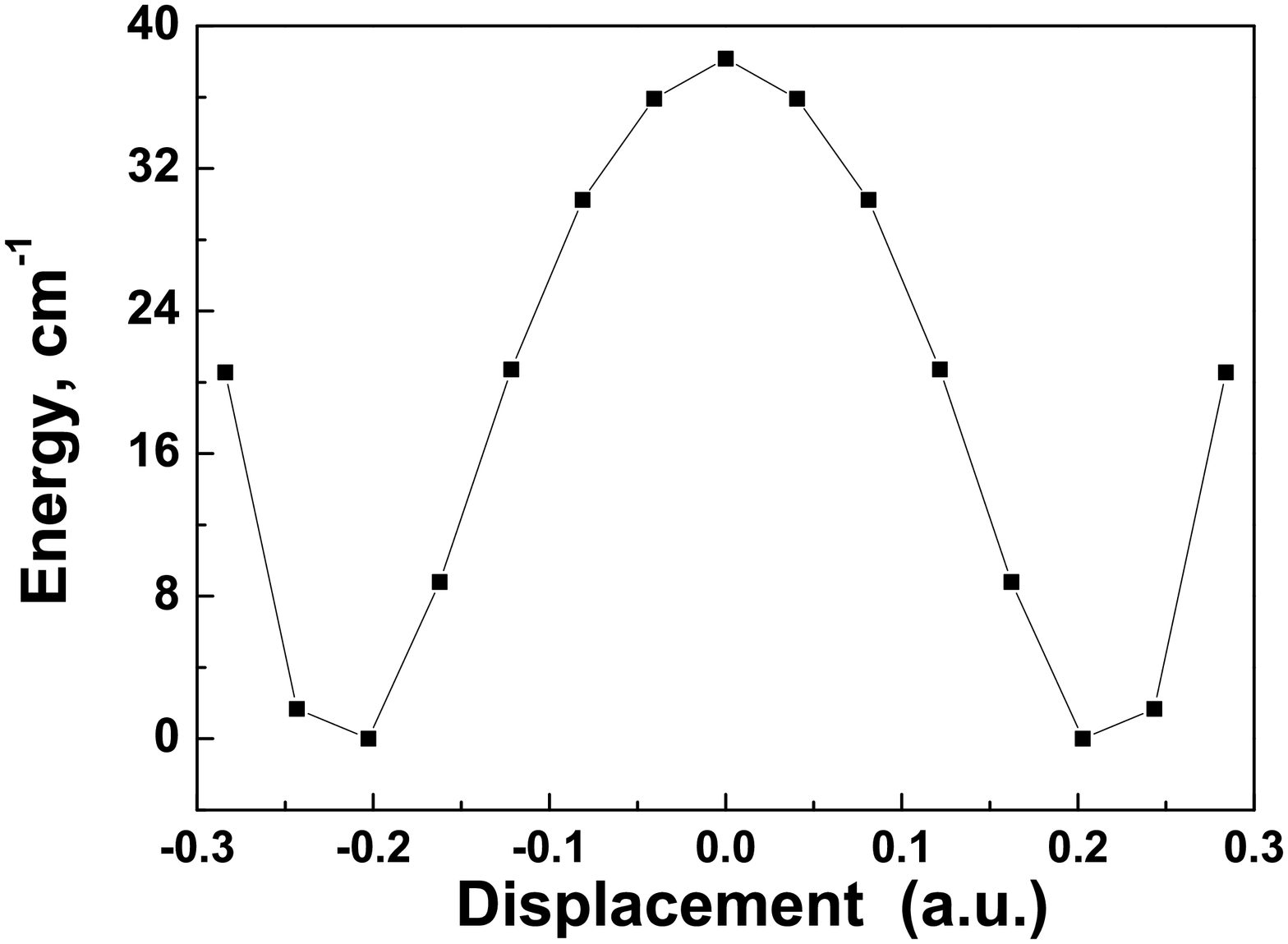}
\caption{Double well potential for the unstable phonon mode using
the frozen phonon method from the calculation with SOC.}
\end{figure}

\begin{figure}[tbp]
\includegraphics [height=3in] {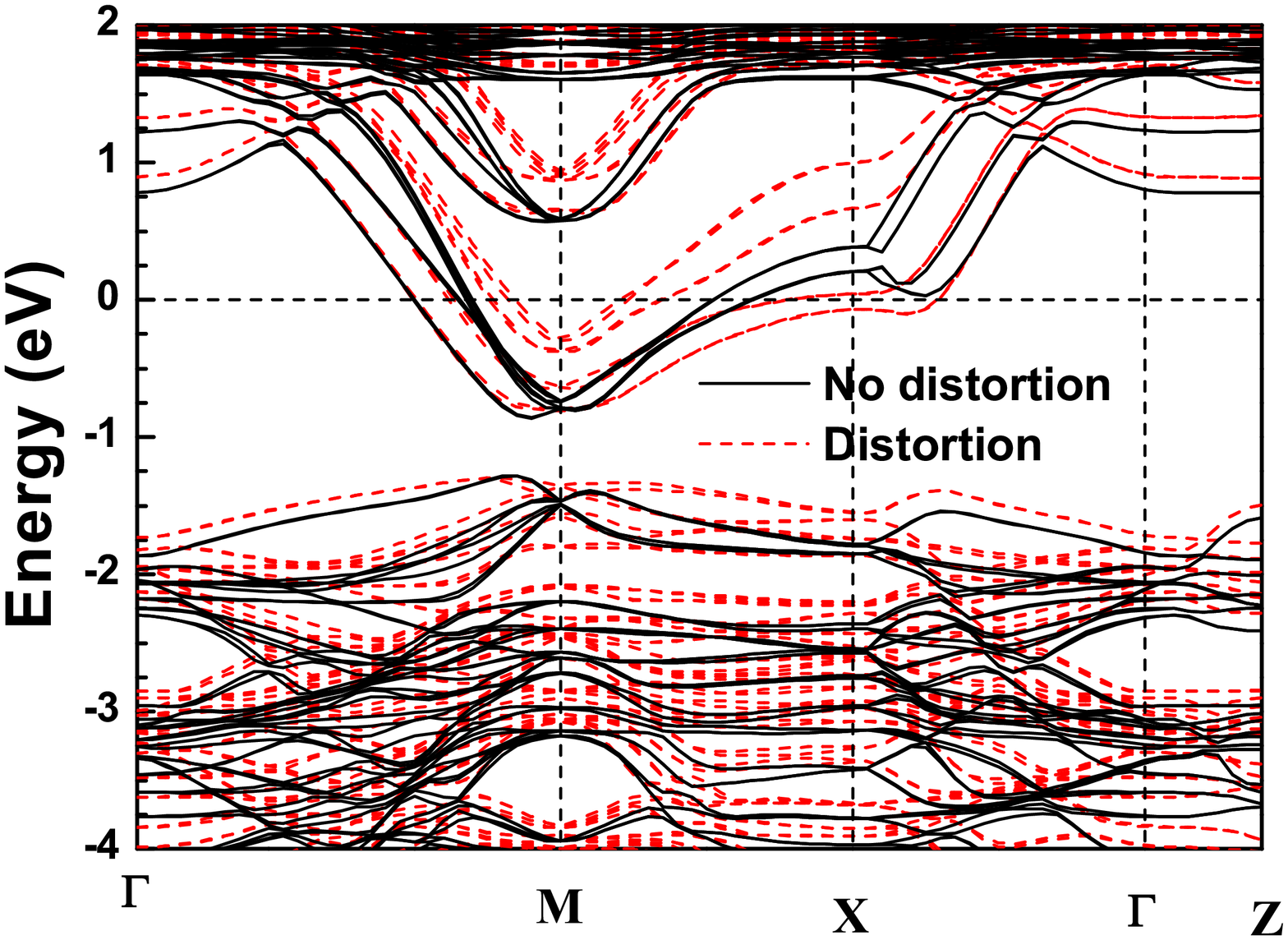}
\caption{Band structure from calculation with SOC. Black line is
the undistorted bands, red line is a distorted bands corresponding
to the \ S1 in-plane displacement.}
\end{figure}

\begin{figure}[tbp]
\includegraphics [height=3in] {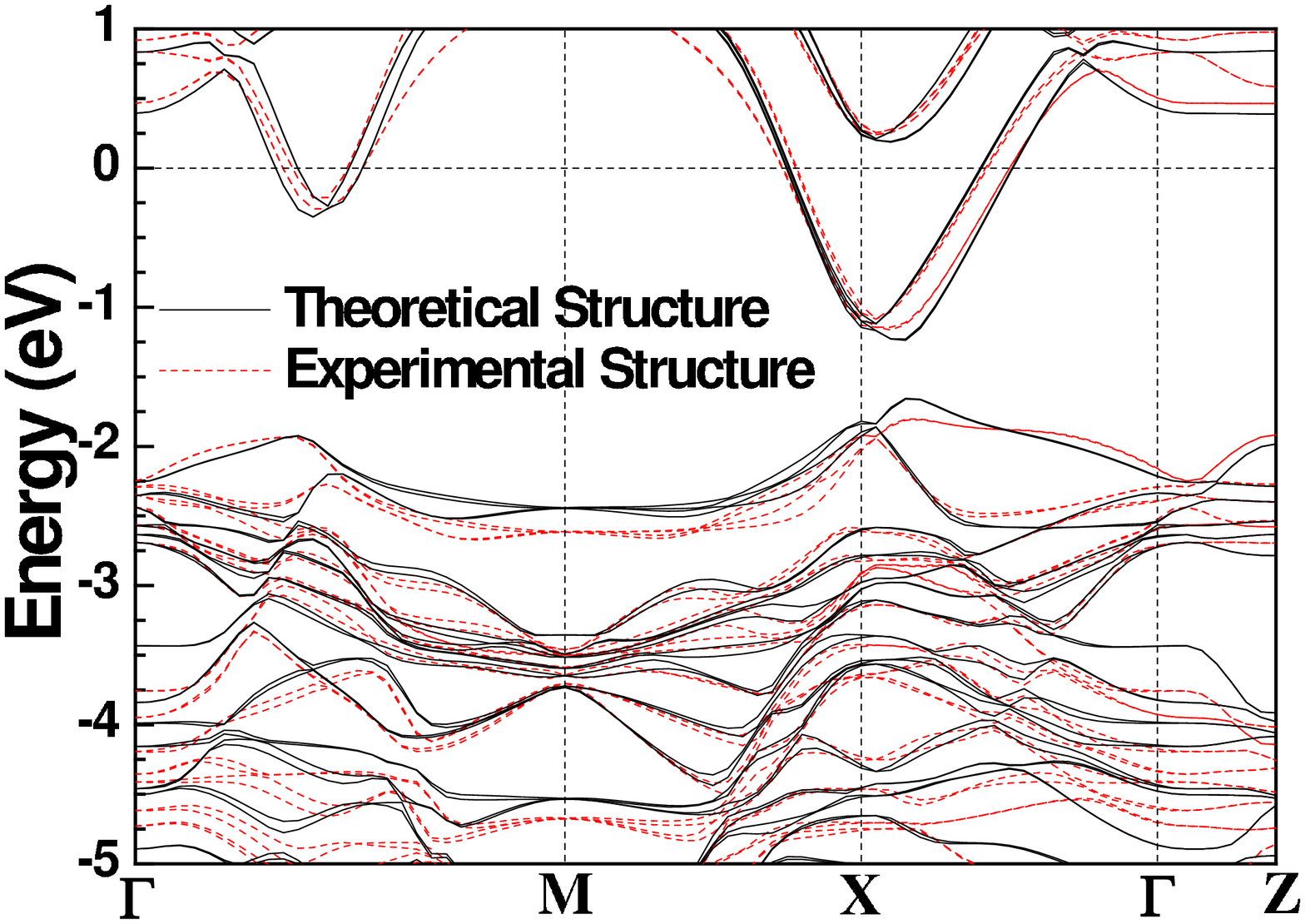}
\caption{Band structure from the calculation with SOC.}
\end{figure}


\begin{thebibliography}{99}
\bibitem{HTC} W. E. Pickett, Rev. Mod. Phys. \textbf{61}, 433 (1989).

\bibitem{Sr2RuO4} Y. Maeno, H. Hashimoto, K. Yoshida, S. Nishizaki, T.
Fujita, J. G. Bednorz and F. Lichtenberg, Nature \textbf{372}, 532 (1994).

\bibitem{MgB2} J. Nagamatsu, N. Nakagawa, T. Muranaka, Y. Zenitani, and J.
Akimitsu, Nature \textbf{410}, 63 (2001).

\bibitem{Fe-based} Y. Kamihara, T. Watanabe, M. Hirano, and Hideo Hosono, J.
Am. Chem. Soc. \textbf{130}, 3296 (2008).

\bibitem{Fe-based-2} J. Paglione and R. L. Greene, Nature Phys. \textbf{6},
645 (2010).

\bibitem{SDW} J. Dong, H. J. Zhang, G. Xu, Z. Li, G. Li, W. Z. Hu, D. Wu, G.
F. Chen, X. Dai, J. L. Luo, Z. Fang and N. L. Wang, Europhys. Lett. \textbf{%
83}, 27006 (2008).

\bibitem{Bi4O4S3} Y. Mizuguchi, H. Fujihisa, Y. Gotoh, K. Suzuki, H. Usui,
K. Kuroki, S. Demura, Y. Takano, H. Izawa and O. Miura, arXiv:1207.3145
(2012).

\bibitem{LaOFBiS2} Y. Mizuguchi, S. Demura, K. Deguchi, Y. Takano, H.
Fujihisa, Y. Gotoh, H. Izawa and O. Miura, arXiv:1207.3558 (2010).

\bibitem{NdOBiS2} S. Demura, Y. Mizuguchi, K. Deguchi, H. Okazaki, H. Hara,
T. Watanabe, S. J. Denholme, M. Fujioka, T. Ozaki, H. Fujihisa, Y. Gotoh, O.
Miura, T. Yamaguchi, H. Takeya and Y. Takano, arXiv:1207.5248 (2012).

\bibitem{Bi4O4S3-2} S. Li, H. Yang, J. Tao, X. Ding and H.H. Wen,
arXiv1207.4955 (2012).

\bibitem{Bi4O4S3-3} S. G. Tan, L. J. Li, Y. Liu, P. Tong, B. C. Zhao, W. J.
Lu, Y. P. Sun, arXiv:1207.5395.

\bibitem{Pressure} H. Kotegawa, Y. Tomita, H. Tou, H. Izawa, Y. Mizuguchi,
O. Miura, S. Demura, K. Deguchi and Y. Takano, arXiv:1207.6935 (2012).

\bibitem{Minimal electronic models} H. Usui, K. Suzuki and K. Kuroki,
arXiv1207.3888 (2012).

\bibitem{SpinExc} T. Zhou and Z. D. Wang, arXiv:1208.1101 (2012).

\bibitem{LaO0.5F0.5BiS2} V. P. S. Awana, Anuj Kumar, Rajveer Jha, Shiva
Kumar, Jagdish Kumar, Anand Pal, Shruti, J. Saha, S. Patnaik,
arXiv:1207.6845 (2012).

\bibitem{Singh} S. K. Singh, A. Kumar, B. Gahtori, Shruti, G. Sharma, S.
Patnaik, V. P. S. Awana, arXiv:1207.5428 (2012).

\bibitem{GGA} J. P. Perdew, K.Burke, and M. Ernzerhof, Phys. Rev. Lett. 77,
3865 (1996).

%\bibitem{the ultrasoft pseudopotential} D. Vanderbilt, Phys. Rev. B \textbf{%
%41}, 7892 (1990).

\bibitem{VASP} G. Kresse and J. Furthmuller, Comput. Mater. Sci. \textbf{6},
15 (1996); G. Kresse and J. Furthmuller, Phys. Rev. B \textbf{54}, 11169
(1996).

\bibitem{pwscf} P. Giannozzi \textit{et al.}, J. Phys.: Condens. Matter,
\textbf{21}, 395502 (2009).

\bibitem{Linear-response} S. Baroni, S. de Gironcoli, A. Dal Corso, and P.
Giannozzi, Rev. Mod. Phys. \textbf{73}, 515 (2001).

\bibitem{Yin-Pickett} Z. P. Yin, S. Lebegue, M. J. Han, B. P. Neal, S. Y.
Savrasov, and W. E. Pickett, Phys. Rev. Lett. \textbf{101}, 047001 (2008).

\bibitem{BaBiO3} V. Meregalli and S. Y. Savrasov, Phys. Rev. B \textbf{57},
14453 (1998).

\bibitem{phonon-linewidth} P. B. Allen, Phys. Rev. B \textbf{6}, 2577 (1972).

\bibitem{linear-response-2} S.Y. Savrasov, Phys. Rev. Lett. \textbf{69},
2819 (1992); S.Y. Savrasov, D.Y. Savrasov, and O.K. Andersen, Phys. Rev.
Lett. \textbf{72}, 372 (1994).

\bibitem{Hui} J.C.K. Hui, and P.B. Allen, J. Phys. F: Met. Phys. \textbf{4},
L42 (1974).

\bibitem{Meregalli} V. Meregalli and S. Y. Savrasov, Phys. Rev. \textbf{57},
14453 (1998).
\end{thebibliography}

\begin{thebibliography}{9}
\bibitem{LaOFBiS2} Y. Mizuguchi, S. Demura, K. Deguchi, Y. Takano, H.
Fujihisa, Y. Gotoh, H. Izawa and O. Miura, arXiv:1207.3558 (2012).

\bibitem{Pb} S. Y. Savrasov and D. Y. Savrasov, Phys. Rev. B 54, 16487
(1996).
\end{thebibliography}
\end{document}